\documentclass[a4paper]{jpconf}
\usepackage{graphicx}
\usepackage{amsmath,amsfonts}
\usepackage{citesort}
\begin{document}

\title{Non-linear parameter estimation for the LTP experiment: analysis of an operational exercise}

\author{G~Congedo$^1$,
        F~De~Marchi$^1$, L~Ferraioli$^1$, M~Hueller$^1$, S~Vitale$^1$,
        M~Hewitson$^2$, M~Nofrarias$^2$, A~Monsky$^2$, M~Armano$^3$,
        A~Grynagier$^4$, M~Diaz-Aguilo$^5$, E~Plagnol$^6$, B~Rais$^6$}

\address{$^1$ Dipartimento di Fisica, Universit\`a di Trento and INFN,
              Gruppo Collegato di Trento, 38123 Povo, Trento, Italy \\
         $^2$ Albert-Einstein-Institut, Max-Planck-Institut f\"ur
              Gravitationsphysik und Universit\"at Hannover, 30167 Hannover, Germany \\
         $^3$ European Space Astronomy Centre, European Space Agency, Villanueva de la
              Ca\~{n}ada, 28692 Madrid, Spain \\
         $^4$ Institut f\"ur Flugmechanik und Flugregelung, 70569
              Stuttgart, Germany \\
         $^5$ UPC/IEEC, EPSC, Esteve Terrades 5,
              E-08860 Castelldefels, Barcelona, Spain \\
         $^6$ APC UMR7164, Universit\'e Paris Diderot, Paris, France}

\ead{congedo@science.unitn.it}

\begin{abstract}
The precursor ESA mission LISA-Pathfinder, to be flown in 2013, aims at demonstrating the feasibility of the free-fall, necessary for LISA, the upcoming space-born gravitational wave observatory. LISA Technology Package (LTP) is planned to carry out a number of experiments, whose main targets are to identify and measure the disturbances on each test-mass, in order to reach an unprecedented low-level residual force noise. To fulfill this plan, it is then necessary to correctly design, set-up and optimize the experiments to be performed on-flight and do a full system parameter estimation. Here we describe the progress on the non-linear analysis using the methods developed in the framework of the \textit{LTPDA Toolbox}, an object-oriented MATLAB Data Analysis environment: the effort is to identify the critical parameters and remove the degeneracy by properly combining the results of different experiments coming from a closed-loop system like LTP.
\end{abstract}

\section{Introduction}

LISA \cite{bender1998} will be the first gravitational wave detector in space. Together with the development of the flight hardware, like the Gravitational Reference Sensors (GRSs) and the interferometers, data analysis plays a crucial role. In order to prove the feasibility of the free-fall, a scientific mission, LISA-Pathfinder (LPF) \cite{vitale2002,armano2009}, has been developed. To date, the mission is in the final stage before the launch: all the subsystems are being fully tested and mostly assembled \cite{mcnamara2008}; at the same time, data analysis is carrying on \cite{antonucci2010,hewitson2009,monsky2009}.

On board LPF, the LTP experiment \cite{anza2005} will verify the free-fall with a unprecedented low-level in force disturbances. In order to achieve this, a measurement scheme has been proposed: two Au-Pt 2-kg Test-Masses (TMs) separated by $\sim$30 cm reproduce one Doppler-link among the six within LISA. A laser interferometer keeps track of the relative displacements and rotations between one TM and the spacecraft and between the two TMs. Also, two GRSs provide the locking of the TMs along the non-optically sensed axes. The aim is to measure the residual forces or any systematics on the first TM, subtract them from the data and improve the quality of the free-fall. A key technique consists on the drag-free: the TM is let in free-fall and the spacecraft is forced to follow the TM itself, based on a suitable readout. At the same time, the second TM is forced to follow the first, in an analogous scheme. The requested forces/torques on the spacecraft and the second TM are then calculated by the Drag-Free Attitude and Control System (DFACS).

In order to measure the force disturbances on the TMs, it is necessary to implement a set of identification experiments, whose main targets are to extract information on the system critical parameters. Each experiment is specifically designed to excite the system and hence its parameters at proper frequencies. The modeled system output, compared to the data, can give estimates for the parameters, but unfortunately some of them will be degenerate. However, by using all the system outputs at the same time, the information is maximized and the parameter degeneracy can be removed: in the end the system is fully identified.

In this paper it is shown how to perform a non-linear time-domain parameter estimation of mission-like simulator data during an operational exercise. In Section \ref{sect:model} we describe the physical model; in Section \ref{sect:technique} we introduce the analysis technique that will be applied subsequently to mission-like simulator data in Section \ref{sect:analysis}. Final comments are discussed Section \ref{sect:final}.

\section{\label{sect:model}LTP as a closed-loop physical system}

As introduced previously, the main LTP science mode consists on one TM in drag-free along the laser-sensitive axis ($x$-axis) and suspended by electrostatic actuators along the other degrees of freedom. The DFACS, based on the interferometer read-out, commands the spacecraft to follow the first TM and the second TM to follow the first.

The 1-D dynamics for the two TMs and the spacecraft (SC) can be easily written in frequency domain as
\begin{align}\label{eq:eom}
    m_1 \, s^2 x_1 & = F_1 - m_1 \, \Gamma_{x,\,2} \left(x_2-x_1\right) - m_1 \, \omega_1^2 x_1 - m_1 \, s^2 x_{\mathrm{SC}}~, \nonumber \\
    \begin{split}
    m_2 \, s^2 x_2 & = F_2 + m_2 \, \Gamma_{x,\,1} \left(x_2-x_1\right) - m_2 \, \omega_2^2 x_2 - m_2 \, s^2 x_{\mathrm{SC}} \\
    & \quad + m_2 \, A_{\mathrm{sus}} H_{\mathrm{sus}} o_{12}~, \nonumber
    \end{split} \\
    \begin{split}
    m_{\mathrm{SC}} \, s^2 x_{\mathrm{SC}} & = F_{\mathrm{SC}} + m_1 \, \omega_1^2 x_1 + m_2 \, \omega_2^2 x_2 - F_1 - F_2 \\
    & \quad - m_2 \, A_{\mathrm{sus}} H_{\mathrm{sus}} o_{12} + m_{\mathrm{SC}} \, A_{\mathrm{df}} H_{\mathrm{df}} o_1~,
    \end{split}
\end{align}

where $s$ is the Laplace frequency and $x_1$ and $x_2$ are the two TMs' coordinates with respect to a local non-inertial reference frame, fixed to the spacecraft. In Eq.\;\eqref{eq:eom}, $F_1$, $F_2$ and $F_\mathrm{SC}$ are all the external forces on both TMs and spacecraft; $m_1 \, \omega_1^2 x_1$ and $m_2 \, \omega_2^2 x_2$ are the residual coupling modeled as oscillator-like forces, being $\omega_1^2$ and $\omega_2^2$ named the \textit{parasitic stiffnesses}; $\Gamma_{x,\,1}$ and $\Gamma_{x,\,2}$ are the mutual gravity gradients between the TMs. The first loop term is $m_{\mathrm{SC}} \, A_{\mathrm{df}} H_{\mathrm{df}} o_1$ which is the DFACS commanded force actuated on the spacecraft by the micro-propulsion thrusters, based on $o_1$, the interferometer reading of $x_1$. The second loop term is $m_2 \, A_{\mathrm{sus}} H_{\mathrm{sus}} o_{12}$, which is the DFACS commanded force actuated on the second TM by the electrostatic suspensions, based on $o_{12}$, the interferometer reading of $x_2-x_1$. Moreover, the loop terms are split into the DFACS control laws, $H_{\mathrm{df}}$ and $H_{\mathrm{sus}}$, and the actual application of the commanded forces, $A_{\mathrm{df}}$ and $A_{\mathrm{sus}}$

Since the optically measured coordinates are $x_1$ and $x_{12}=x_2-x_1$, whilst $x_{\mathrm{SC}}$ is not directly detected, it is natural to eliminate the latter via elementary algebra and define the new physical variable $\mathbf{x}=\left(\begin{smallmatrix}x_1\\x_{12}\end{smallmatrix}\right)$. Provided that:
\begin{description}
  \item[$o_1$\;:] the interferometer read-out of the displacement $x_1$ of the spacecraft to the first TM;
  \item[$o_{12}$\;:] the interferometer read-out of the displacement $x_{12}$ of the second TM to the first TM;
\end{description}
then $\mathbf{x}$ converts to the interferometer read-outs through $\mathbf{o}=\mathbf{S}\cdot\mathbf{x}+\mathbf{o}_n$, where $\mathbf{o}=\left(\begin{smallmatrix}o_1\\o_{12}\end{smallmatrix}\right)$, $\mathbf{o}_n$ are the interferometer read-out noises and $\mathbf{S}$ is the interferometer conversion matrix containing possible cross-coupling terms as off-diagonal terms. Hence, the first two equations of motion in Eq.\;\eqref{eq:eom} can be rearranged in a single matrix equation and, by solving with respect to $\mathbf{o}$, it is possible to demonstrate that $\mathbf{o} = \mathbf{o}_s + \mathbf{o}_n$, which contains a deterministic part $\mathbf{o}_s$ and a noise contribution $\mathbf{o}_n$ from the read-out and force noises. The deterministic part is essentially made of injected stimuli to the system both in the controller setpoint (interferometer guidances) and as direct applied forces. In this paper we refer to the analysis of only the first case. Consequently, if $\mathbf{o}_i$ are the inputs to the controller setpoint, the deterministic part is given by $\mathbf{o}_s = \mathbf{H}_\mathrm{mdl} \cdot \mathbf{o}_i$, where $\mathbf{H}_\mathrm{mdl}$ is the transfer function matrix model from the input guidances to the interferometer outputs.

\section{\label{sect:technique}Parameter estimation technique}

Once derived the deterministic part $\mathbf{o}_s$, it is logical to assert that exciting the system in one channel and analyzing both outputs, the system can be fully characterized and information on the critical parameters can be extracted. Finally, this knowledge takes to the exact computation of the residual force noise on each TM \cite{ferraioli2010b}, which is a key target of the mission.

Recall now what is assumed to be known: firstly, the interferometer transfer function matrix obtained from the equation of motion, say $\mathbf{H}_\mathrm{mdl}$, which depends on $\mathbf{p}$, the set of all critical parameters (see Section \ref{sect:parameters}); in the end, the interferometers injections $\mathbf{o}_i$. Therefore, the modeled outputs to be compared to the interferometer read-outs $\mathbf{o}_\mathrm{data}$ in time domain are
\begin{equation}\label{eq:mdl}
\mathbf{o}_{\mathrm{mdl}}(t,\mathbf{p}) = \mathfrak{F}^{-1}_t \left[\mathbf{H}_\mathrm{mdl}(\omega,\mathbf{p}) \cdot \mathbf{o}_i(\omega)\right]~,
\end{equation}
where $\mathfrak{F}^{-1}_t$ is the inverse Fourier operator and $\omega$ the Fourier frequency. The comparison between data and model is performed on whitened time-series in a least-square sense. The whitening technique is part of a more general framework, the colored noise generation \cite{ferraioli2010a}.

As a brief summary, the estimation:
\begin{enumerate}
\item assumes \textit{gaussianity} and \textit{stationarity};
\item treats whitened data and models;
\item works in time domain;
\item is non-linear since $\mathbf{H}_\mathrm{mdl}$ in Eq.\;\eqref{eq:mdl} is so with respect to each parameter;
\item is a multi-experiment/multi-channel analysis.
\end{enumerate}

\section{\label{sect:analysis}Analysis of operational exercise data}

The aim of this section is to apply the concepts described in Section \ref{sect:model} and \ref{sect:technique} to simulated data. The LISA-Pathfinder STOC Simulator (LSS) is a software written for the ESA-STOC center by the aerospace industry to fulfill the requirements of very realistic mission-like data and test all the methods developed by the analysts. Recently, it has been proposed to simulate a number of exercises and practice on these data sets. Here the system identification of an operational exercise has been performed in a very mission-like manner. The final goal is to give an estimate of the best-fit parameters and their uncertainties, in a similar way it will be done during the mission.

\subsection{\label{sect:parameters}System parameters}

Here is a description of the system critical parameters being considered in the analysis:
\begin{description}
  \item[$H_\mathrm{df}$\;:] drag-free gain;
  \item[$H_\mathrm{sus}$\;:] electrostatic suspension gain;
  \item[$S_{21}$\;:] interferometer sensing matrix coefficient, named \textit{cross-coupling} or \textit{cross-talk}, from $o_{i\,1}$ and $o_{12}$;
  \item[$\omega_1^2$\;:] stiffness of the first TM;
  \item[$\omega_2^2$\;:] stiffness of the second TM;
  \item[$\Delta t_\mathrm{loop,\,1}$\;:] time-delay in the control loop for the first read-out;
  \item[$\Delta t_\mathrm{loop,\,12}$\;:] time-delay in the control loop for the second read-out;
  \item[$\tau_1$\;:] characteristic time for the thruster response;
  \item[$\tau_2$\;:] characteristic time for the electrostatic suspension response.
\end{description}

\subsection{Stating the experiments}

When defining the type of identification experiments to be performed during the mission, there is a general freedom in selecting the injections to be applied to the system. However, the system is resolved and the information maximized if the proper frequencies are input: a series of sine waves makes the role of the actual stimulus to the system being studied.

Two general classes of experiments are described: the \textit{unmatched-} and \textit{matched-} stiffness. In the second case, the control mode is set up so that the stiffnesses are artificially matched to the same value. In this paper we focus only to the unmatched-stiffness case and, in order to estimate the physical parameters, two types of identification experiments can be thought. Therefore we refer to:
\begin{description}
\item Exp.\;1 when injecting $o_{i,\,1}$ into the interferometer guidance for $o_{1}$;
\item Exp.\;2 when injecting $o_{i,\,12}$ into the interferometer guidance for $o_{12}$.
\end{description}

\begin{figure}[htb]
\centering
\begin{tabular}{cc}
\hspace{-10pt}\includegraphics[height=120pt]{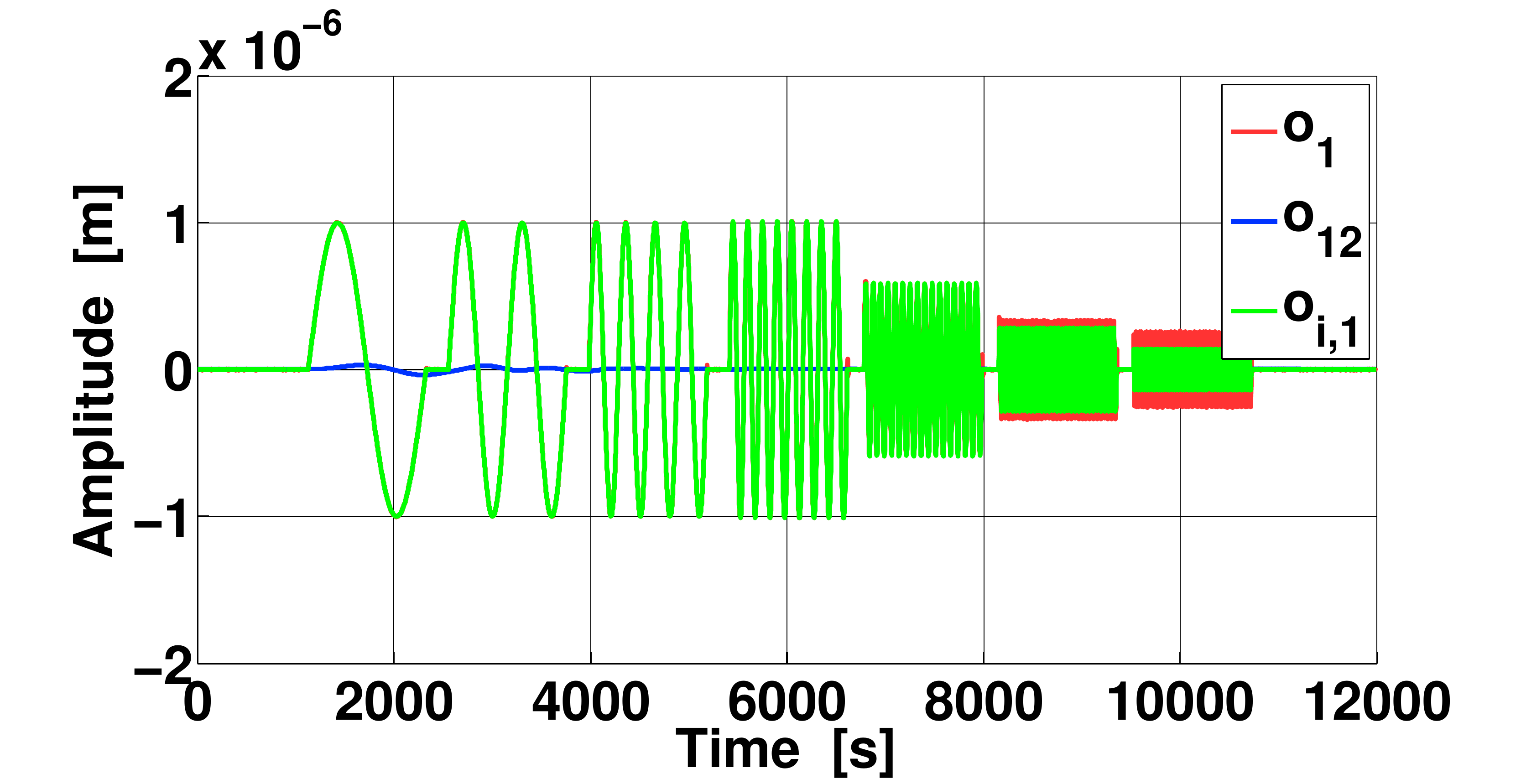} &
\hspace{-10pt}\includegraphics[height=120pt]{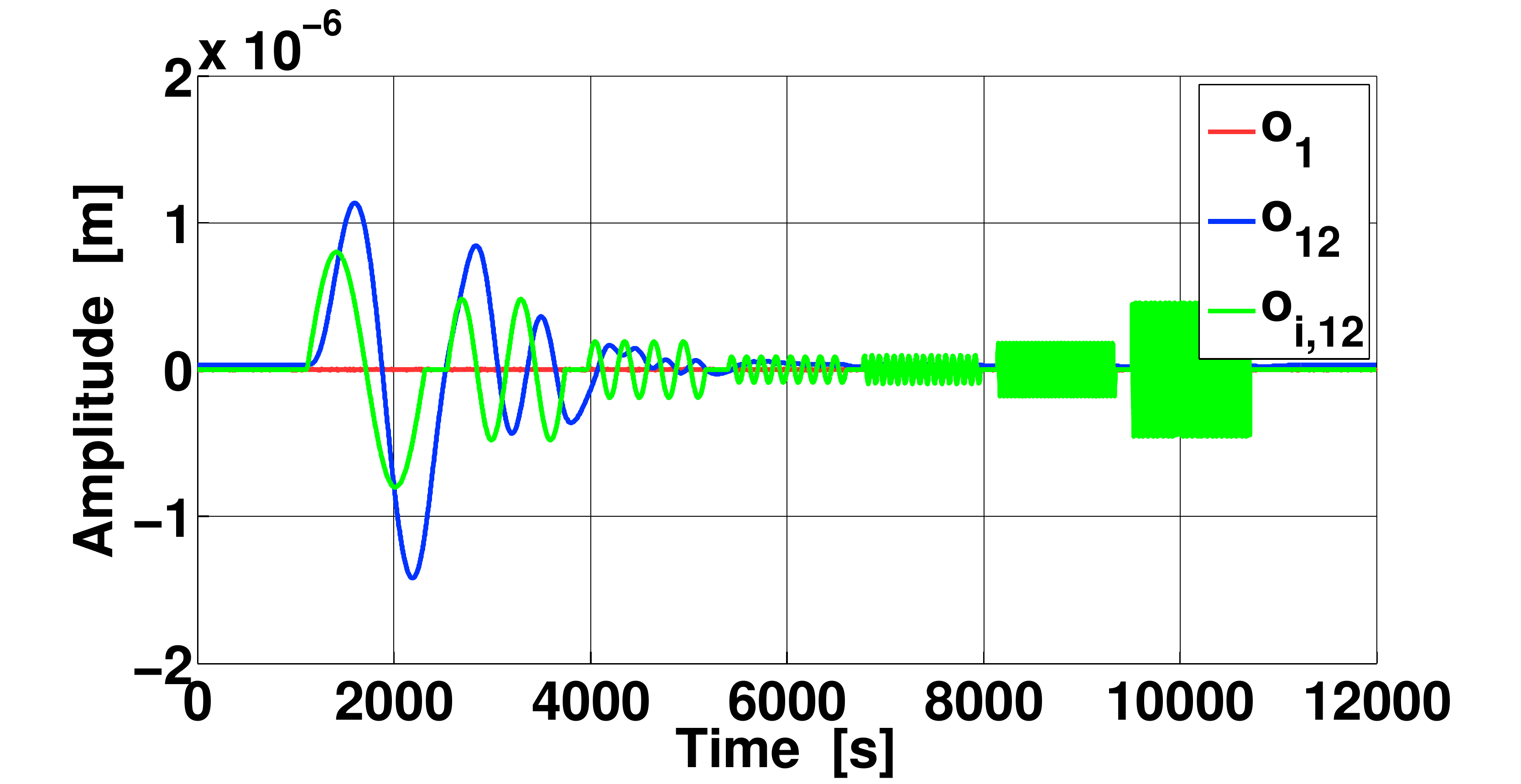} \\
(a) & (b)
\end{tabular}
\caption{\label{fig:o1_o12_oi}\footnotesize{Interferometer outputs $o_1$, $o_{12}$ and guidances for (a) Exp.\;1 ($o_{i,\,1}$) and (b) Exp.\;2 ($o_{i,\,12}$).}}
\end{figure}

For the sake of clarity, in Figure \ref{fig:o1_o12_oi} the interferometer outputs and guidances are shown for both experiments. The inputs that define the two injection experiments are made of a series of sine waves of pre-determined frequencies selected to excite specifical elements of the transfer function matrix.

\subsection{Estimation}

The parameter estimation is performed by following the ideas in Section \ref{sect:technique} and the modeled outputs are computed through Eq.\;\eqref{eq:mdl}. The whitening filters are computed on long noise data stretches before the injection and applied both to the modeled outputs at each iteration step and to data.  The estimation is focused on a joint analysis of all four read-outs coming from both experiments: this maximizes the information and allows the identification of all parameters. For a detailed description on non-linear estimation methods and their issues, see for example \cite{press2007}. The algorithm has been implemented in the \textit{LTPDA Toolbox} \cite{ltpda} under the MATLAB programming interface \cite{matlab}.

Final results are contained in Table \ref{tab:best-fit}. As it is clear, except for $H_\mathrm{sus}$ and $S_{21}$, whose best-fit is compatible (within the error) to the respective initial guess, all the others are moved well apart. The errors are computed in the linear approximation, hence only analytical first derivatives of the modeled output are taken into account.
\begin{table}[htb]
\caption{\label{tab:best-fit}\footnotesize{Best-fit parameters at $\chi^2/\nu=1.2$ and $\nu=21964$.}}
\begin{center}
\begin{tabular}{llll}
\br
Parameter                           & Guess              & Best-fit              & 1-$\sigma$ error \\
\mr
$H_\mathrm{df}$                     & 1                  & 1.07988               & $6.5\cdot10^{-4}$ \\
$H_\mathrm{sus}$                    & 1                  & 0.999998              & $3.9\cdot10^{-5}$ \\
$S_{21}$                            & 0                  & $-1.4\cdot10^{-6}$    & $1.2\cdot10^{-6}$ \\
$\omega_1^2\,[s^{-2}]$              & $-1.3\cdot10^{-6}$ & $-1.3207\cdot10^{-6}$ & $2.6\cdot10^{-9}$ \\
$\omega_2^2\,[s^{-2}]$              & $-1.4\cdot10^{-6}$ & $-2.0356\cdot10^{-6}$ & $2.5\cdot10^{-9}$ \\
$\Delta t_\mathrm{loop,\,1}\,[s]$   & 0.3                & 0.19875               & $4.8\cdot10^{-4}$ \\
$\Delta t_\mathrm{loop,\,12}\,[s]$  & 0.3                & 0.2048                & $8.0\cdot10^{-3}$ \\
$\tau_1\,[s]$                       & 0.1                & 0.4109                & $1.9\cdot10^{-3}$ \\
$\tau_2\,[s]$                       & 0.01               & 0.1984                & $2.8\cdot10^{-3}$ \\
\br
\end{tabular}
\end{center}
\end{table}
The best-fit values for this data set are also confirmed by other independent estimation methods developed within the LTPDA team, so we are confident on the numbers proposed in Table \ref{tab:best-fit}.

\begin{figure}[htb]
\centering
\begin{tabular}{cc}
\hspace{-10pt}\includegraphics[height=120pt]{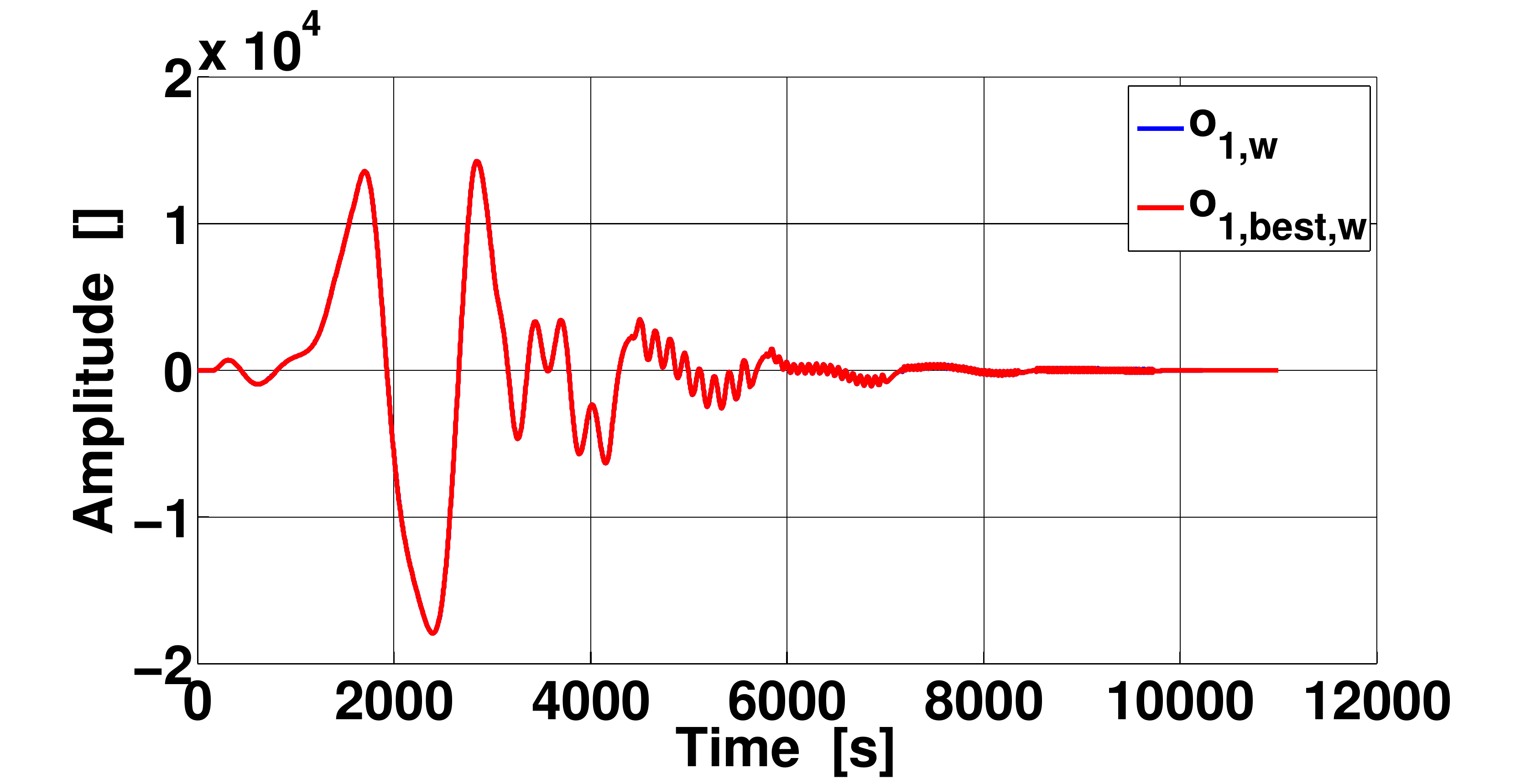} &
\hspace{-10pt}\includegraphics[height=120pt]{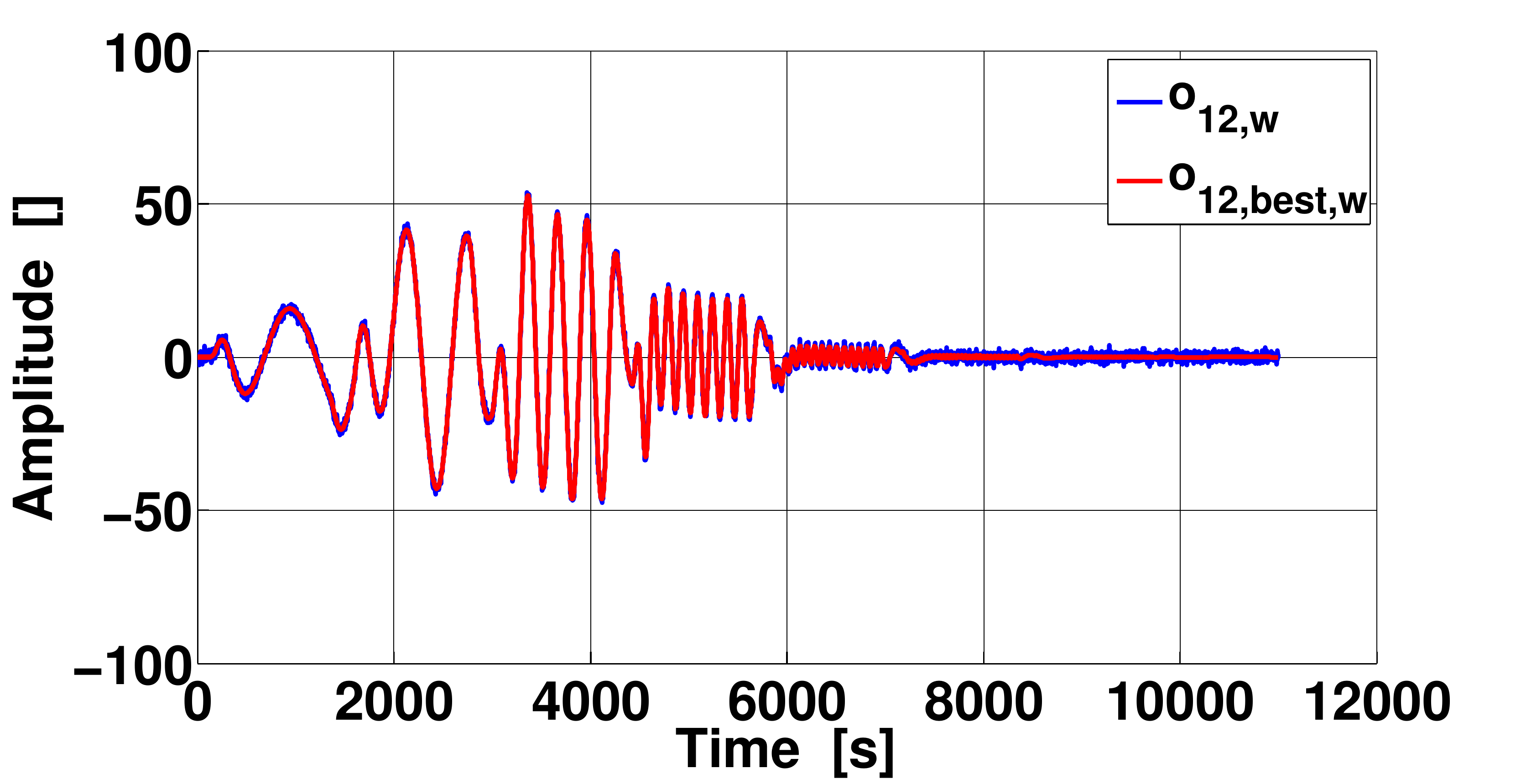} \\
(a) & (b) \\
\hspace{-10pt}\includegraphics[height=120pt]{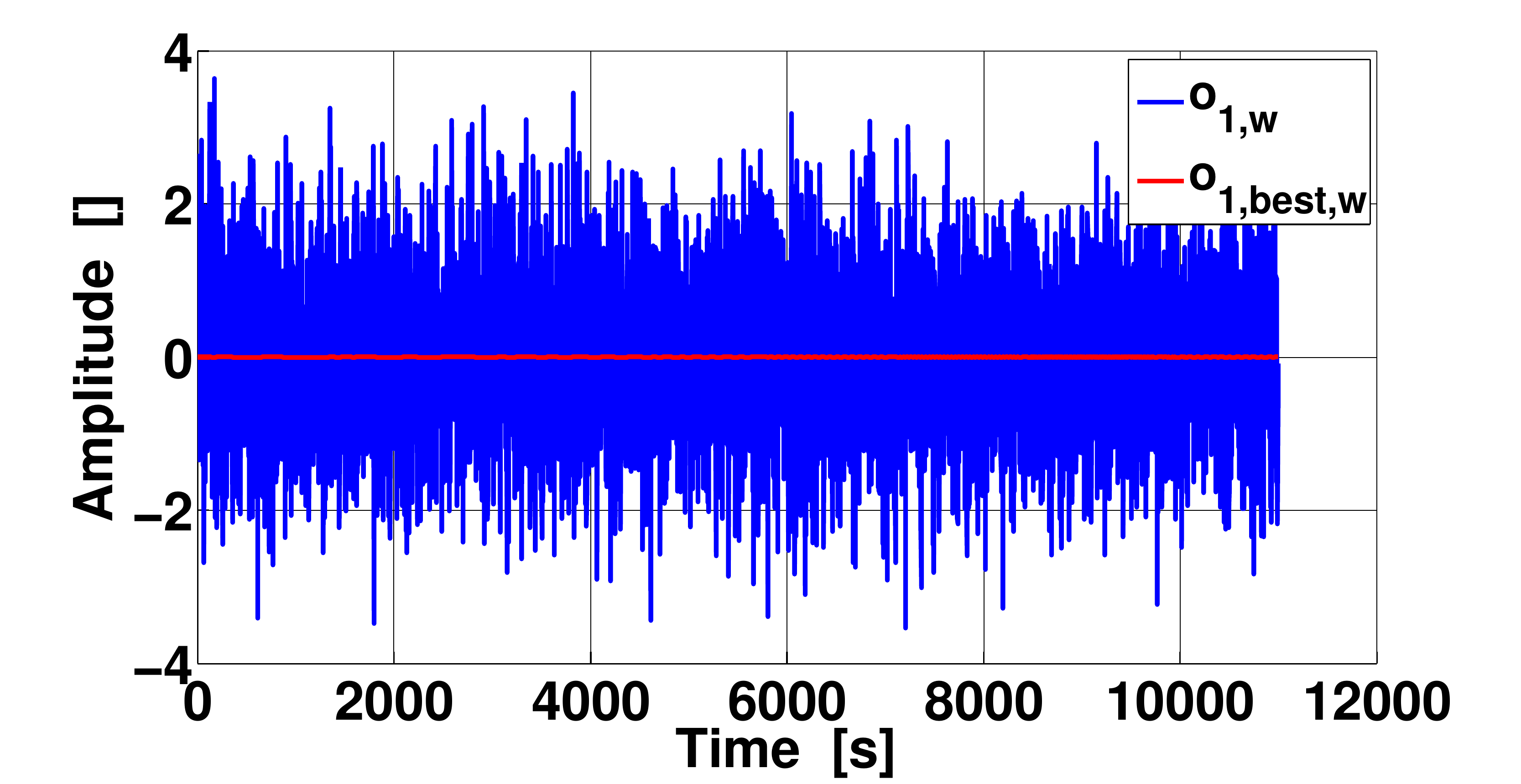} &
\hspace{-10pt}\includegraphics[height=120pt]{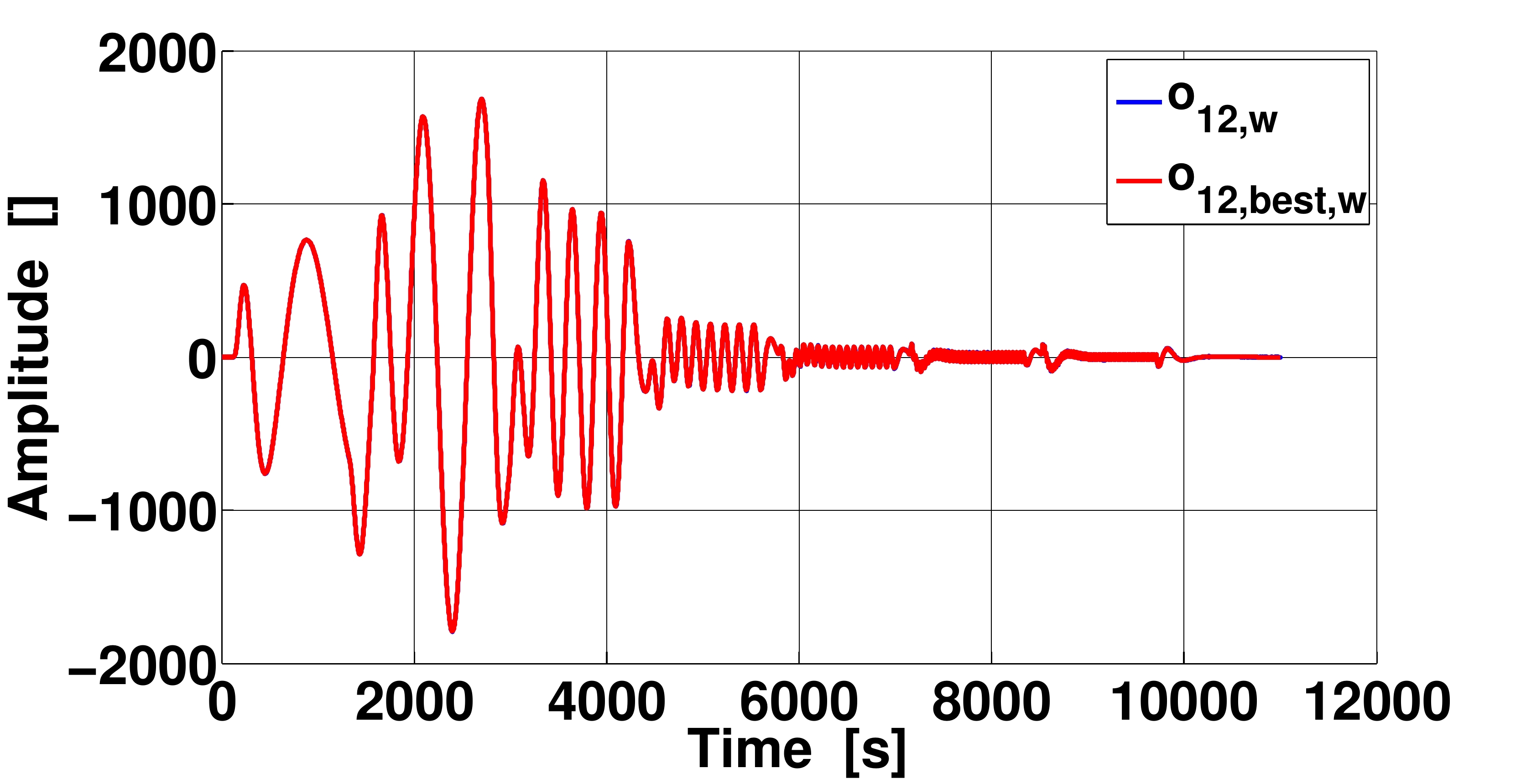} \\
(c) & (d)
\end{tabular}
\caption{\label{fig:best-fit}\footnotesize{Whitened best-fit model and data for: (a) Exp.\;1 first read-out, (b) Exp.\;1 second read-out, (c) Exp.\;2 first read-out, (d) Exp.\;2 second read-out.}}
\end{figure}

Figure \ref{fig:best-fit} shows the plots of the best-fit model compared to the respective read-out, both whitened, (panels (a) and (b)) for Exp.\;1 and Exp.\;2, respectively. The agreement is also confirmed by the fact that the final residuals are compatible with a zero-mean gaussian white noise with unitary standard deviation. In particular, for Exp.\;1 the agreement between data and model is acceptable expect for the first read-out (see panel (a) in \ref{fig:best-fit}) where some systematics still remain, but of the order of 1$\%$. Instead, for Exp.\;2 it is clear that the first read-out (see panel (c) in \ref{fig:best-fit}) contains no signal: indeed there is no cross-coupling $S_{12}$ between the two read-outs: that is why we have not taken into account in the analysis.

As a final consideration, it is important to point out that the approach to parameter estimation in this paper is clearly experimental: no one exactly knows which are the real parameters. Anyway, some practical considerations and the agreement with other methods suggest that in principle the method is capable to extract the system critical parameters, among those, the most important are surely the two stiffnesses and the interferometer cross-coupling.

\section{\label{sect:final}Final remarks}

The main scientific payload onboard LPF, the LTP experiment, will demonstrate that is possible to keep a TM in free-fall with an unprecedented low level in force disturbances. After the completion of the LPF mission, this expertise will be transferred to LISA, the gravitational wave observatory in space. The LTPDA team is working in developing all the data analysis procedures to characterize and study the system. For it, parameter estimation has a key role and in this paper a method has been proposed to be used during the mission.

The proposed method is a non-linear, time-domain, least-square minimization of a joint $\chi^2$ that embraces the parameter information from all experiments and read-outs. The overall complication is twofold. From one side, it is easy to write the dynamics in frequency domain but its impossible to derive analytical models in time domain. From the other side, the LTP noise shape is not flat for both read-outs and the whitening filters are derived to properly whiten data and models. In the end, the whitening allows us to treat data in least-square sense and compute both the best-fit parameters and their covariance matrix.

The STOC simulator has given the opportunity to use and test the method in the same way as it could be done during the LPF mission. The non-linear estimation technique has shown appreciable results evident both in the plots of the best-fit curves and the analysis of the residuals, but also in agreement with other independent methods.

Further investigations will surely come up on new STOC data with increasing realism, until the mission will start in the next years. For the moment, the time-domain non-linear method seems to be one of the most promising parameter estimation solution to be considered for the mission.

\appendix

\section*{References}
\bibliography{bibliography}

\end{document}